\documentclass[aps,showpacs,pra]{revtex4}

\usepackage[normalem]{ulem}
\usepackage{exscale}
\usepackage{graphicx}
\usepackage{amsmath}
\usepackage{latexsym}
\usepackage{amsfonts}
\usepackage{amssymb}
\usepackage{times}

\usepackage{epsfig,pstricks}
\providecommand{\NN}[0]{\mathbb{N}}

\def\DJo{$\;$\kern-.4em \hbox{D\kern-.8em\raise.15ex\hbox{--}\kern.35em okovi\'c}}
\def\CC{{\rm\kern.24em \vrule width.04em height1.46ex depth-.07ex
\kern-.30em C}}
\def\P{{\rm I\kern-.25em P}}
\def\NN{{\rm I\kern-.25em N}}
\def\RR{{\rm
         \vrule width.04em height1.58ex depth-.0ex
         \kern-.04em R}}
\def\id{{\rm 1\kern-.22em l}}
\def\ZZ{{\sf Z\kern-.44em Z}}

\def\trace{{\rm tr}\;}

\newtheorem{psatz}{Satz}[section]
\newtheorem{pdef}{Definition}[section]
\newtheorem{conjecture}{Vermutung}[section]
\newtheorem{Theorem}{Theorem}[section]
\newtheorem{Corollary}{Corollary}[section]

\newenvironment{since}{${}^{}$\\ {\bf Proof}:  \begin{quote}\begin{em}}{\begin{flushright}
                   {\bf q.e.d.}\end{flushright}\end{em}\end{quote} }

\newenvironment{eqblock}[2]{\beq\label{#2}\begin{array}{#1}}{\end{array}
                                \eeq}
\newenvironment{neqblock}[1]{\[\begin{array}{#1}}{\end{array}\]}
\newcommand{\beqb}{\begin{eqblock}}
\newcommand{\eeqb}{\end{eqblock}} 
\newcommand{\nbeqb}{\begin{neqblock}}
\newcommand{\neeqb}{\end{neqblock}}

\newcommand{\beq}{\begin{equation}}
\newcommand{\beqa}{\begin{eqnarray}}
\newcommand{\eeq}{\end{equation}}
\newcommand{\eeqa}{\end{eqnarray}}
\newcommand{\nbeqa}{\begin{eqnarray*}}
\newcommand{\neeqa}{\end{eqnarray*}}

\newcommand{\ket}[1]{| #1 \rangle}

\newcommand{\Matrix}[2]{\left( \begin{array}{#1} #2 \end{array}
  \right)}

\begin{document}

\title{Classification scheme of pure multipartite states based on topological phases}

\author{Markus Johansson$^{1}$, Marie Ericsson$^2$, Erik Sj\"oqvist$^{1,2}$, and Andreas Osterloh$^{3}$}
\affiliation{$^1$Centre for Quantum Technologies, National University
of Singapore, 3 Science Drive 2, 117543 Singapore, Singapore }
\affiliation{$^2$Department of Quantum Chemistry, Uppsala University, Box 518, 
SE-751 20 Uppsala, Sweden}
\affiliation{$^3$Fakult{\"a}t f{\"u}r Physik, Campus Duisburg, Universit{\"a}t Duisburg-Essen, Lotharstr. 1, 47048 Duisburg, Germany.}

\date{\today}

\begin{abstract}
We investigate the connection between the concept of affine balancedness (a-balancedness) introduced 
in [Phys. Rev A. {\bf 85}, 032112 (2012)] and polynomial local SU invariants 
and the appearance of topological phases respectively. It is found that different 
types of a-balancedness correspond to different types of local SU invariants 
analogously to how different types of balancedness as defined in  
[New J. Phys. {\bf 12}, 075025 (2010)] correspond to different types of local 
SL invariants. These different types of SU invariants distinguish between states 
exhibiting different topological phases. In the case of three qubits the different 
kinds of topological phases are fully distinguished by the three-tangle together 
with one more invariant.
Using this we present a qualitative classification scheme based on balancedness 
of a state. 
While balancedness and local SL invariants of bidegree $(2n,0)$ classify the 
SL-semistable states [New J. Phys. {\bf 12}, 075025 (2010), 
Phys. Rev. A {\bf 83} 052330 (2011)], a-balancedness and local SU invariants 
of bidegree $(2n-m,m)$ gives a more fine grained classification. 
In this scheme the a-balanced states form a bridge from the genuine entanglement 
of balanced states, invariant under the SL-group, towards the entanglement of 
unbalanced states characterized by U invariants of bidegree $(n,n)$.
As a by-product we obtain generalizations to the $W$ state, i.e., states that 
are entangled, but contain only globally distributed entanglement of 
parts of the system. 

\end{abstract}


\maketitle

\section{Introduction}

One element of quantum mechanics that appears counter intuitive is 
certainly entanglement.
Though it is present everywhere where 
there is an interaction, its effects are most easily observed at low temperature and in 
carefully controlled environments. But even at ambient conditions the extent to which entanglement plays a role in what we observe in nature is discussed
(see e.g.
\cite{Fleming10,CaiBriegel10,CarusoPlenio10,Cai2010,TierschBriegel12}).
Local SU invariance is a feature of any entanglement property and thus enters 
as a minimal requirement for every entanglement measure~\cite{MONOTONES}.
Therefore, it has been used for the classification of
states \cite{LindenPopescu98,GrasslRoetteler98,CarteretLinden99,LindenPopescu99} 
and the definition of entanglement monotones as measures
\cite{ParkerRijmen01,wei03,Scott04,BuchMintert04,Mintert05,Demkowicz06}.
The group of local operations which leaves entanglement properties invariant was recognized rather 
early to be the special linear (SL) group 
\cite{Duer00,Wong00,VerstraeteDM02,VerstraeteDM03,JaegerII03}, which is the group 
underlying the stochastic local operations and classical communication (SLOCC).
However, the classification of multipartite entanglement is difficult and, 
due to the fact that almost all polynomial entanglement measures 
have a polynomial degree of at least four \footnote{For measures quadratic
in $\psi$, see Refs. \cite{Wootters98,Uhlmann00}.}, 
their mixed state extension via the convex roof is also problematic, although 
solvable for certain cases~\cite{LOSU,EOSU,Jung09,HigherRanktau3,Eltschka2012,Siewert2012}. 
  
The classification based on polynomial SL and unitary invariants is 
complete in the sense that it fully distinguishes inequivalently 
entangled states from each other. However, one can also consider 
other classification schemes that are less distinguishing, i.e., more 
coarse grained, but which focus on qualitative properties of the states. 
Such classification schemes can group sets of SLOCC classes or 
alternatively local SU(2) classes into families based on some 
common property \cite{VerstraeteDMV02,Lamata07,Bastin09,Viehmann}.

The qualitative property we will consider here, is balancedness. 
The concept of balancedness as defined in \cite{OS09} is a property 
related to genuine entanglement which is taken here to mean entanglement for which there are measures constructed from polynomial SLOCC invariants \cite{Coffman00,OS09}. 
This relation between balancedness 
and polynomial invariants is described by the notion of semistability in geometric invariant theory 
\cite{Mumford,Dolgachev}. 
Recently, it has also been shown in \cite{Johansson12} that balancedness 
is a useful concept for describing the appearance of topological 
phases \cite{MilmanMosseri03,Milman06,Du07,Souza07,Oxman11,Johansson12,Khoury13,JohanssonKhoury13}.
The analysis of the ${\rm{SU(2)}}^{\otimes q}$ topological phases  
of pure $q$-qubit states \cite{Johansson12} has generalized the 
balancedness condition \cite{OS09} relevant for
local SL(2) invariance~\cite{Luque02,Brylinski02a,Brylinski02b,VerstraeteDM03,
Luque05,OS04,OS05,DoOs08} 
to an affine balancedness, or a-balancedness, relevant for local SU(2) invariance.
In this analysis, a splitting of the three-qubit
$W$ SLOCC class was found, since the a-balanced state 
$\ket{W'}:=\frac{1}{2}(\ket{000}+\ket{100}+\ket{010}+\ket{001})$, is distinguished by a topological phase $\pi$ from the unbalanced $W$ state
$\ket{W}:=\frac{1}{\sqrt{3}} (\ket{100}+\ket{010}+\ket{001})$, which has none.
This raises the question of whether this qualitative difference between these two SLOCC 
equivalent states is captured by a local SU(2) invariant that distinguishes 
the states~\cite{Johansson12}.

In this paper we show that there indeed is a local SU(2) invariant 
that captures this difference between states in the SLOCC $W$ class with different 
topological phases.
Furthermore, we show that for an arbitrary number of qubits, different types of 
a-balanced states correspond to different local SU(2) invariant polynomials and 
these capture the qualitative properties related to topological phases. 
This is analogous to how the original balancedness condition 
is related to local SL(2) invariant polynomials and 
topological phases. In this light, we also discuss how 
multipartite entangled states can be classified based on their 
balancedness properties.

The paper is organized as follows. In the next section, we briefly 
discuss balancedness and its relation to topological phases. 
Section \ref{3bit} is devoted to the three 
qubit case and in particular the $W'$ state, where we write down the 
SU(2) invariant of bidegree $(3,1)$ that detects it (i.e, assigns a nonzero value to it). 
It is then followed by Sec. \ref{gen}, which is an investigation 
of the general case of SU(2) invariants of bidegrees $(2n-p,p)$ 
with $p\leq n$. 
In Sec. \ref{ind}, we discuss how a-balancedness can be used 
for a classification of multipartite entanglement.
In the last section, we draw our conclusions, and give an outlook.

\section{Balanced states}\label{bal}
The first observation of the relation between three-partite entanglement and the property later termed balancedness in \cite{OS09} was reported by Coffman, Kundu, 
and Wootters in \cite{Coffman00}. 
This concept has been found to be far more general, 
and has been extended and applied to genuinely multipartite
entangled states in \cite{OS09}.
We briefly give the main definition of a {\em balanced state}
here. Suppose we have a $q$-qubit state with a decomposition using $L$ basis state 
vectors of a computational basis, called its length. 
Then, we construct a $q\times L$ matrix $A_{kl}$ such that
each column of $A_{kl}$ corresponds to one of these basis state vectors in such a way that the 
$k$th element 
of the $l$th column is the $k$th entry of the $l$th basis state vector,
but with every $0$ replaced by $-1$.
A state is balanced {\em iff} positive numbers in $\NN$ exist such that

\beq
n_l A_{kl}=0\ ; \quad n_l \in\NN\ \forall \ k=1,\dots,q.
\eeq

A state is called partly balanced if some of the numbers must be chosen to be zero.
Furthermore, a balanced state is called {\em irreducible} if no part of the 
matrix is balanced by itself \cite{OS09}.
As an example, the $q$-qubit GHZ state
$\alpha \ket{0\dots 0} + \beta\ket{1\dots 1}$ would be given by
\beq
A_{kl}=\Matrix{cc}{
 -1 & 1\\
\vdots & \vdots\\
-1 & 1}
\eeq 

A state that has a balanced part in every computational product basis is genuinely entangled.
The irreducible balanced states are particular in that they indicate a basis 
of entangled states
detected by measures derived from SL invariants and are not separable over any bipartition. 
Therefore, they are genuinely multipartite entangled~\footnote{We mention here, that the 
terminus ``genuinely multipartite entangled'' is widely used for saying, that a state is not bipartite 
in the literature. See e.g. ~\cite{Duer00,Bourennane04}. We use the more strict 
terminology from ~\cite{OS04} and call such states ``globally entangled''.}.
Furthermore, balancedness in every computational basis implies that the possible 
total phase factors that 
can be accumulated in a cyclic evolution generated by local SU(2) or SL(2) evolutions is 
a discrete set \cite{Johansson12}. 
This discretization of the possible accumulated phases is related to the 
nontrivial topology of the local SU(2) or SL(2) orbits, and the phases are therefore 
called topological.

The study of topological phases generated by local SU(2) operations prompted the 
extension of the concept of balancedness. In this extension the requirement that 
the numbers $n_{l}$ in the definition of balancedness belong to the natural numbers 
$\NN$ is replaced by the requirement that they are nonzero integers, i.e., they belong to 
$\ZZ\backslash{0}$.
The states balanced in $\NN$ have been termed convexly balanced or c-balanced, while 
the states balanced in $\ZZ\backslash{0}$ have been termed affinely balanced or a-balanced 
in \cite{Johansson12}.

Let us next focus on the a-balanced states. Those of the a-balanced states that 
are not also c-balanced, do not have genuine entanglement,
hence all local SL-invariant measures vanish, but 
if they are a-balanced in every basis, then they have globally distributed entanglement.
Therefore, consider a state which is irreducibly a-balanced in one (local) basis,
\beq
z_l A_{kl}=0\ ; \quad z_l\in\ZZ\setminus \{0\}\ \forall \ k=1,\dots,q
\eeq
where $q$ is the number of qubits and $l=1,\dots , L$, where $L$ is the length of the state. 
What changes in the 
definition, when compared to c-balancedness, 
is that negative numbers are also admitted, such that we
are in $\ZZ$ and none of the $z_l$ are zero. 
Therefore, some of the proofs of \cite{OS09}
on c-balanced states are also valid for a-balanced states. We therefore have the following
\begin{Theorem}
Product states are never irreducibly a-balanced.
\end{Theorem}  
\begin{Theorem}
Every a-balanced $q$-qubit state with length larger than $q+1$ is reducible.
\end{Theorem}

Every c-balanced state is mapped into an a-balanced state
if some of its components in the computational basis decomposition are being 
spin flipped ($\sigma_y^{\otimes q}C$), without producing a product state. 
As we will show, each such spin flip corresponds to a map from an invariant 
detecting the c-balanced state to an invariant detecting the a-balanced state. 
This map takes invariants into invariants of a different bidegree. 
As an example, the three-tangle of bidegree $(4,0)$ can be mapped to an SU invariant 
of bidegree $(3,1)$, and then further to a U invariant of bidegree $(2,2)$. 
We have to consider, at most, half of the components being
spin flipped (any multiplicity of a component in the balancedness relation is counted),
since a state with $n\geq L/2$ could be obtained from the spin flipped version of 
the c-balanced state with only $k=L-n\leq L/2$ components flipped.

A consequence of irreducible balancedness of a state, which we will make use of, is that 
the multiset of integers $z_l$ is uniquely defined up to a common factor. Thus, there 
is a unique multiset $\{z_{0},z_{1},\dots,z_{L-1}\}$, up to a factor of $-1$, of integers 
without a common prime factor associated to an irreducible state.
Furthermore, we will sometimes consider sets of states that are balanced in the same way, 
in the sense that there is a particular matrix $A_{kl}$ such that each state in the set is local 
unitary equivalent to a form where its balancedness is described by $A_{kl}$. Such a set 
will be referred to as an $A$-class of states.

\section{Three qubits}
\label{3bit}

Before dealing with an arbitrary number of qubits, we consider the irreducible states, 
topological phases, and invariants of three qubits since this is a well understood case \cite{Duer00,Carteret00b,Acin00,Carteret00}.
Here the primary entangled states are the globally entangled $W$ state,
$\ket{W}=(\ket{100}+\ket{010}+\ket{001})/\sqrt{3}$, and the genuinely three-party 
entangled GHZ state, $\ket{GHZ}=(\ket{111}+\ket{000})/\sqrt{2}$~\cite{Duer00}.
The latter has an equivalent representation $\ket{GHZ}\stackrel{SU(2)}{\cong} 
\ket{X}:=(\ket{111}+\ket{100}+\ket{010}+\ket{001})/2$, which would be the generalization of the 
$X$ state in \cite{OS05}. Moreover, 
$\ket{\widetilde{GHZ}}:=(\alpha\ket{111}+\beta \ket{000})/\sqrt{2}
\stackrel{SL(2)}{\cong} \ket{\widetilde{X}}:= (a \ket{111}+ b \ket{100} + c \ket{010} + d\ket{001})/2$
for general non-vanishing parameters, following \cite{OS09}. 
However, it is easy to check that
$\ket{\widetilde{GHZ}}\stackrel{SU(2)}{\not\cong}
\ket{\widetilde{X}}$. 
The states $\ket{\widetilde{GHZ}}$ and 
$\ket{\widetilde{X}}$ 
are  distinguished by topological phases since 
$\ket{\widetilde{X}}$ 
has topological phases that are multiples of $\frac{\pi}{2}$ 
while $\ket{\widetilde{GHZ}}$ only has the phase $\pi$ unless 
$|\alpha|=|\beta|=\frac{1}{\sqrt{2}}$.

For the $W$ state, the situation is analogous. Here, 
$\ket{W}\stackrel{SL(2)}{\cong}\ket{W'}:=
(\ket{000}+\ket{100}+\ket{010}+\ket{001})/2$ \cite{Duer00}, 
whereas 
$\ket{W}\stackrel{SU(2)}{\not\cong}\ket{W'}$. 
The latter is an a-balanced state \cite{Johansson12} and has a topological phase
of $\pi$, whereas the $W$ state is completely unbalanced and has none.
Now, if there was a nonvanishing SU(2) invariant of bidegree 
$(d_1,d_2)$, where $d_{1}\neq{d_{2}}$, 
it has been noted in \cite{Johansson12} that a topological phase 
$\chi = \frac{2m\pi}{d_1-d_2}$ for some integer $m$ would appear.
It would therefore be promising, if an SU(2)-invariant of a bidegree that 
corresponds to a $\pi$ phase exists that 
is nonzero for $\ket{W'}$.

It has been described in \cite{Luque06} how to construct local SU(2) invariants. 
For three qubits the algebra of local SU-invariants has seven primary generators and three 
secondary generators \cite{Grassl02}.
The primary generators contain one invariant of bidegree $(4,0)$ as well as its 
complex conjugate of bidegree $(0,4)$, 
one of bidegree $(3,1)$ and its complex conjugate of bidegree $(1,3)$, 
one of bidegree $(1,1)$ (the squared modulus of the state), 
and two of bidegree $(2,2)$ (these correspond to the two independent reduced density matrices). 
For our purposes, we identify invariants with their complex conjugates. 
The $(4,0)$ invariant is the three-tangle $\tau_{(4,0)}=\tau_3$~\cite{Coffman00}. 
The invariant of bidegree $(3,1)$ has been calculated from \cite{Luque06} to be
\beqa\label{Invar31}
\tau_{3,1}&=&\sum_{i_1,i_2=0}^1(|\psi_{0i_1i_2}|^2-|\psi_{1i_1i_2}|^2)
(\psi_{000}\psi_{111}+\psi_{100}\psi_{011}-\psi_{010}\psi_{101}+\psi_{001}\psi_{110})\nonumber \\
&+& 2(\psi_{0i_1i_2}(\psi_{100}\psi_{111}-\psi_{110}\psi_{101})\psi^*_{1i_1i_2}
-\psi_{1i_1i_2}(\psi_{000}\psi_{011}-\psi_{010}\psi_{001})\psi^*_{0i_1i_2})\ ,
\eeqa
where $\psi_{i_{1}i_{2}i_{3}}$ are the coefficients of the state vector in the computational basis.
It indeed detects $\ket{W'}$ and not $\ket{W}$, and is manifestly SU(2) invariant. 
This explains the fact that the $W'$ state has a topological 
phase of $\chi=\frac{2m \pi}{d_1-d_2}= m \pi$ for $m\in \NN$, whereas the $W$ state has none. 
The invariant also detects $\ket{\widetilde{GHZ}}$ but not $\ket{\widetilde{X}}$, which is only detected by the threetangle, and this explains 
the different topological phases of these states.

From these observations we can draw the full picture of polynomial SU invariants 
of bidegree $(d_{1},d_{2})$, where $d_{1}\neq{d_{2}}$, and 
topological phases for three qubits. 
As shown by Ac\'{\i}n {\it et al.} \cite{Acin00}  and Carteret {\it et al.} \cite{Carteret00}, 
each three-qubit state can be transformed 
by local unitary operations to a canonical form. In particular we consider the canonical form where the set of basis vectors is invariant under permutation of the qubits \cite{Carteret00}

\begin{eqnarray}
\kappa_{0}e^{i\theta}|000\rangle+\kappa_{1}|001\rangle+\kappa_{2}|010\rangle
+\kappa_{3}|100\rangle+\kappa_{4}|111\rangle,
\end{eqnarray}
where $\kappa_{j}$ for $j=0,1,2,3,4,$ and $\theta$ are real numbers. For a generic three qubit state all $\kappa_{j}$ in the canonical form are nonzero.
Thus, a generic entangled three-qubit state can be transformed 
by local unitaries to a balanced but not irreducibly balanced form. 
The states that can be transformed to an irreducible form are subsets characterized by fewer parameters.
There 
are three different A-classes
of irreducible states.
The first we will call the $X$-class, with reference to ~\cite{OS05}, which is given by $\kappa_{0}=0$ and the other coefficients are nonzero; 
the second is the $\widetilde{GHZ}$ class given by 
$\kappa_{1}=\kappa_{2}=\kappa_{3}=0$ while $\kappa_{4}$ and $\kappa_0$ are nonzero,
and the $W'$ class is given by $\kappa_{4}=0$.
Thus, the set of local SU(2) orbits belonging to the $X$-class and the set 
belonging to the $W'$ class are both four-parameter subsets, 
while the SU(2) orbits of the $\widetilde{GHZ}$ class are a two-parameter set. The $\widetilde{GHZ}$ class intersects 
with the $X$ class in the SU(2) orbit of the GHZ state.

The $X$ class is detected by the three-tangle $\tau_3$, but not by $\tau_{3,1}$. 
The $W'$ class on the other hand, is detected by $\tau_{3,1}$, 
but not by $\tau_3$. While the  modulus of $\tau_{3}$ has its unique maximum 
on the local unitary orbit of the $X$ state, the modulus of $\tau_{3,1}$ has 
its unique maximum on the local unitary orbit of $W'$. 
All genuinely entangled states except the $X$ class are detected by both 
$\tau_3$ and $\tau_{3,1}$. 
For three qubits there is thus a clear relation between the irreducible 
balanced states, the polynomial SU invariants of bidegree $(d_1,d_{2})$, 
where $d_{1}\neq{d_{2}}$, and topological phases. An experimental proposal
to observe the topological phases in three-qubit systems was recently 
given in \cite{JohanssonKhoury13}.

\section{The general case: more than three qubits}\label{gen} 

In this section, we show that for every type of irreducible a-balancedness, there is 
at least one local SU-invariant polynomial that detects it - i.e., it gives a nonzero 
value to it - while not detecting other types of irreducible a-balancedness. 
Furthermore, if such an ${\rm{SU(2)}}^{\otimes{q}}$-invariant polynomial of 
bidegree $(d_{1},d_{2})$, where 
$d_{1}\neq{d_{2}}$, detects a $q$-qubit state, then this state exhibits topological 
phases~\cite{Johansson12}.
For three qubits, we have seen that the irreducible a-balanced state 
$|W'\rangle$ is detected by an SU(2) invariant polynomial of this kind.
Now we show that every irreducible a-balanced state 
for which the sum of the corresponding integers $\{z_{0},z_{1},\dots,z_{L-1}\}$ is nonzero has topological phases,  
and that there always are SU(2)-invariants of this type that detect them.

In Sects. \ref{secs1} to \ref{secs4}, we will go through the steps leading to this 
conclusion. First, in Sec. \ref{secs1}, we introduce a mapping between irreducible 
c-balanced states and irreducible a-balanced states that are not c-balanced.
We then show in Sec. \ref{secs2} that this mapping between states induces a mapping 
between SL invariants that detect irreducible c-balanced states, and SU invariants that detect 
the a-balanced but not c-balanced states.
Although this induced mapping does not give the explicit form of the SU invariants, 
it allows us to deduce some of their properties in Sec.~\ref{secs4}.

\subsection{The partial spin flip}
\label{secs1}

We introduce a mapping between states.
The mapping makes use of the universal spin-flip transformation $\sigma_{y}^{\otimes{q}}C$, where $C$ is the complex conjugation that acts only on the state vector. This is the unique antiunitary transformation that flips arbitrary spins \cite{Buzek99} since it is invariant under local SU and local SL transformations \cite{OS04}. The spin-flip transformation can be applied to an arbitrary number of qubits and has been used, for example, to construct entanglement measures such as the concurrence \cite{Wootters98}.
Given a $q$-qubit state $|{\psi}\rangle$, we can express it as a sum of two components, 
$|{\psi}\rangle\equiv|\phi\rangle+|\theta\rangle$. We then apply a spin-flip transformation
only to $|\theta\rangle$, which gives

\begin{eqnarray}\label{ett}
|\psi\rangle\to|\tilde{\psi}\rangle&\equiv&|\phi\rangle
+\sigma_{y}^{\otimes{q}}C|\theta\rangle\equiv|\phi\rangle+|\tilde{\theta}\rangle.
\end{eqnarray}
The spin-flip transformation $\sigma_{y}^{\otimes{q}}C$ is almost a conjugation, 
$\sigma_{y}^{\otimes{q}}C\sigma_{y}^{\otimes{q}}C=(-1)^q\id$,
and therefore $\sigma_{y}^{\otimes{q}}C|\tilde{\theta}\rangle\equiv (-1)^q|\theta\rangle$.
The kind of mapping in Eq. (\ref{ett}) will be referred to as a "partial spin flip" map, 
and it is only defined relative to a given decomposition of the state vector 
into two terms. In particular, for a computational basis $\{\ket{ijk\dots}\}$ where $0$ is replaced by $-1$, we have 
\beq
\ket{\theta}=\sum \theta_{ijk\dots}\ket{ijk\dots},\phantom{u}\textrm{and}\phantom{u}\ket{\tilde{\theta}}=\sum \tilde{\theta}_{ijk\dots}\ket{ijk\dots}
\eeq
with
\beq\label{theta-tilde}
\tilde{\theta}_{ijk\dots}=(-i)^{(i + j + k + \dots)} \theta^*_{-i-j-k\dots}\; .
\eeq

Assume that $|\psi\rangle$ is an irreducible c-balanced state of length $L$, where 
the terms are indexed by $\{0,1,\dots,L-1\}$. Choose a decomposition of 
$|\psi\rangle$ such that $|\theta\rangle$ corresponds to a subset $s$ of the terms 
of $|\psi\rangle$ and apply the corresponding partial spin-flip map to produce 
a state $|\tilde{\psi}\rangle$. Then, $|\psi\rangle$ and $|\tilde{\psi}\rangle$ 
can be written as

\begin{eqnarray}\label{tva}
|{\psi}\rangle=\sum_{j=0}^{L-1}\psi_{j}\otimes_{k=1}^{q}|A_{kj}\rangle\nonumber\\
|\tilde{\psi}\rangle=\sum_{j\notin{s}}\psi_{j}\otimes_{k=1}^{q}|A_{kj}\rangle\nonumber\\
+\sum_{j\in{s}}(-i)^{\sum_{k}A_{kj}}\psi^*_{j}\otimes_{k=1}^{q}|-A_{kj}\rangle\nonumber\\
\equiv{\sum_{j}\tilde{\psi}_{j}\otimes_{k=1}^{q}|\tilde{A}_{kj}\rangle},
\end{eqnarray} 
where $\psi_{j}$ and $\tilde{\psi}_{j}$ are the expansion coefficients of 
$|\psi\rangle$ and $|\tilde{\psi}\rangle$, respectively.

We remark that the state $|\tilde{\psi}\rangle$ is not a c-balanced state. 
Depending on the choice 
of $s$, it can be either an irreducible a-balanced state, 
a state where one or more qubits are in a tensor product with 
the remaining qubits, 
or, in some cases, an entangled state without topological phases. 
This will be further 
elaborated in Sec. \ref{secs5}. 
Furthermore, every a-balanced state that is not 
already a c-balanced state can be mapped into a c-balanced state by some partial spin flip.

\subsection{Induced mapping between invariants}
\label{secs2}

\begin{Theorem}
Consider a state $|\psi\rangle$ for which there is a nonvanishing polynomial 
SL-invariant $P$. For every state $|\tilde{\psi}\rangle$ obtained from $|\psi\rangle$ 
through a partial spin flip, there exists an invariant function $\tilde{P}$ 
on the SU(2) orbit of $|\tilde{\psi}\rangle$.
\end{Theorem}

\begin{since}
To see this, consider a c-balanced state $|\psi\rangle$ in an arbitrary local product basis

\begin{eqnarray}|\psi\rangle=\sum{\psi}_{ijk\dots}|ijk\dots\rangle,\end{eqnarray}
where $ijk\dots$ is a string of $1$s and $-1$s.
Assume that there is a ${\rm{SL(2)}}^{\otimes{q}}$-invariant polynomial 
$P$ that detects $|\psi\rangle$. Formally, $P$ can be expressed as

\begin{equation}P=\sum_{\alpha}b_{\alpha}\prod{\psi}_{ijk\dots}^{r(\alpha)_{ijk\dots}},\end{equation}
for some sets of exponents $\{r(\alpha)_{ijk\dots}\}$ and constants $b_{\alpha}$.

Consider then a decomposition of $|\psi\rangle$ as 
$|\psi\rangle=|\phi\rangle+|\theta\rangle$. 
With use of the notation 
$|\phi\rangle=\sum{\phi}_{ijk\dots}|ijk\dots\rangle$, 
$|\theta\rangle=\sum{\theta}_{ijk\dots}|ijk\dots\rangle$, 
the polynomial $P$ can be expressed, using \eqref{theta-tilde}, 
in the variables $\phi_{ijk\dots}$ and $\theta_{ijk\dots}$ as

\begin{equation}
P=\sum_{\alpha}b_{\alpha}\prod(\phi_{ijk\dots}+\theta_{ijk\dots})^{r(\alpha)_{ijk\dots}}\; .
\end{equation}
Consider then the partially spin-flipped state 
$|\tilde{\psi}\rangle=|\phi\rangle+|\tilde{\theta}\rangle$ 
associated to the above decomposition, as given by Eq. (\ref{ett}). 
Using the notation 
$|\tilde{\theta}\rangle=\sum{\tilde{\theta}}_{ijk\dots}|ijk\dots\rangle$ 
where $\theta_{ijk\dots}=i^{({i+j+k+\dots})}\tilde{\theta}^*_{-i-j-k\dots}$, 
we can express $P$ in $\phi_{ijk\dots}$ and $\tilde{\theta}_{-i-j-k\dots}$ as

\begin{equation}\label{sont}
P=\sum_{\alpha}b_{\alpha}\prod(\phi_{ijk\dots}+i^{({i+j+k+\dots})}
\tilde{\theta}^*_{-i-j-k\dots})^{r(\alpha)_{ijk\dots}}
\end{equation}

This second expression for $P$ in Eq. (\ref{sont}) defines a function 
$\tilde{P}$ on the ${\rm{SU(2)}}^{\otimes{q}}$ orbit of $|\tilde{\psi}\rangle$ 
such that $P(|{\psi}\rangle)=\tilde{P}(|\tilde{\psi}\rangle)$.
Furthermore any ${\rm{SU(2)}}^{\otimes{q}}$ operation commutes with the 
spin flip operation $\sigma_{y}^{\otimes{q}}C$. 
Therefore, $P(U|\psi\rangle)=\tilde{P}(U|\tilde{\psi}\rangle)$ 
for $U\in{{\rm{SU(2)}}^{\otimes{q}}}$.
Since $P$ is, in particular, invariant on the ${\rm{SU(2)}}^{\otimes{q}}$ orbit 
of $|{\psi}\rangle$, it follows that $\tilde{P}$ is invariant on the 
${\rm{SU(2)}}^{\otimes{q}}$ orbit of $|\tilde{\psi}\rangle$. 
\end{since}
Note that since only values of homogeneous invariants are concerned, 
$\tilde{P}$ itself does not need to be a homogeneous function. 
Note also that the above argument can be made in the other direction as well. 
That is, the existence of a polynomial SU invariant $\tilde{P}$ 
that detects states on the ${\rm{SU(2)}}^{\otimes{q}}$ orbit of 
$|\tilde{\psi}\rangle$ implies the existence of an SU invariant function 
$P$ that evaluates to a nonzero value on the 
${\rm{SU(2)}}^{\otimes{q}}$ orbit of $|{\psi}\rangle$ related to 
$\tilde{P}$ by the induced mapping.

\subsection{Irreducible c-balanced states and SL invariant polynomials}
\label{secs3}

The irreducibility of a state places a constraint on the homogeneous 
degrees of any polynomial SL invariant that detects the state.
Consider an irreducible $q$-qubit $c$-balanced state $|\psi\rangle$ 
of length $L$

\begin{eqnarray}
|\psi\rangle=\sum_{j=0}^{L-1}\psi_{j}\otimes_{k=1}^{q}|A_{kj}\rangle,
\end{eqnarray}
where $A_{kj}=1,-1$. We assume that the terms are indexed such that the 
multiset of integers $\{z_{0},z_{1},\dots,z_{L-1}\}$ associated with the state, 
where each $z_{j}$ is associated to the term with coefficient $\psi_{j}$, 
and satisfies $|{z}_{0}|\geq|{z}_{1}|\geq|{z}_{2}|\geq\dots\geq|{z}_{L-1}|$. 
Furthermore, we assume that the $z_j$ have no common divisor.
Since the irreducible c-balanced states are genuinely multipartite entangled,
there is a homogeneous ${\rm{SL(2)}}^{\otimes{q}}$-invariant 
polynomial $P$ that evaluates to a nonzero value for this state.

In the particular basis that we have chosen for the state, the monomials of $P$ 
that detect $|\psi\rangle$ are of the form $\sum_{\alpha}\prod_{j}\psi_{j}^{r_{\alpha{j}}}$, 
up to constant factors, for some sets of exponents $\{r_{\alpha{j}}\}$. 
We call the polynomial in the $\psi_{j}$'s made up of these monomial terms $P_A$. 
$P$ and $P_{A}$ evaluate to the same value for every state related to  
$|\psi\rangle$ by local filtering operations.

Consider a particular local filtering operation $F$ on the $k$th qubit 

\begin{eqnarray}F=\left(\begin{array}{cc}
t^{-1} & 0  \\
0 & t  \\
\end{array}\right).
\end{eqnarray}
This operation multiplies the $\alpha$th monomial of $P_{A}$ 
by $t^{\sum_{j}A_{kj}r_{\alpha{j}}}$. Since the polynomial $P$ is 
${\rm{SL(2)}}^{\otimes{q}}$ invariant, the sum of all the monomials 
of $P_{A}$ after the filtering must equal the sum before the filtering. 
Moreover, this must be true for any $t$. This is possible only if 
$t^{\sum_{j}A_{kj}r_{\alpha j}}=1$ for each $\alpha$. Since this must be true for local filterings on any qubit, this gives us a system of $q$ linear equations. 
Using the convention that the term of $|\psi\rangle$ with 
coefficient $\psi_{0}$ is $|11\dots{1}\rangle$, this system of equations 
can be formulated as a matrix equation:

\begin{eqnarray}\label{col}\left(\begin{array}{cccc}
A_{21} & {A}_{31} & \cdots & {A}_{(L-1)1}  \\
A_{22} & {A}_{32} & \cdots & {A}_{(L-1)2} \\
\vdots & \vdots & \ddots &\vdots \\
A_{2q} & {A}_{3q}& \cdots & {A}_{(L-1)q}\\
\end{array}\right)\left(\begin{array}{c}
r_{\alpha{1}}  \\
r_{\alpha{2}}\\
\vdots\\
r_{\alpha{L-1}}\\
\end{array}\right)=-r_{\alpha{0}}\left(\begin{array}{c}
1  \\
1\\
\vdots\\
1\\
\end{array}\right),
\end{eqnarray}
for each $\alpha$. Since the state is irreducible, and thus the columns 
of the matrix on the left hand side are linearly independent, the matrix 
on the left-hand side uniquely determines the solution up to a common 
multiplicative factor $h$. Equation (\ref{col}) is the same equation that 
appeared in \cite{Johansson12} to determine the set $\{z_{0},z_{1},\dots,z_{L-1}\}$. 
The solutions are given by $\{z_{0},z_{1},\dots,z_{L-1}\}$ such that 
$r_{\alpha{j}}=h_{\alpha}z_{j}$ for some $h_{\alpha}$. Since the polynomial is homogeneous,
we have only one $h_{\alpha}\equiv{h}$. 
Thus, the polynomial $P_{A}$ is a single monomial $P_{A}$ given, 
up to a constant, by

\begin{eqnarray}\label{pa}P_{A}=\left (\prod_{j=0}^{L-1}\psi_{j}^{z_{j}}\right )^h.\end{eqnarray}
We will briefly comment on the nature of $h$. As the $\{ z_0,\dots , z_{L}\}$
are assumed to have no common divisor, any solution to Eq. \eqref{col} that
occurs is $h\in \mathbb{N}$, where an $h\geq 2$ means that the length $l=\sum_j z_j$ of the state fits $h$ times in the polynomial degree of the corresponding
invariant~\cite{OS09,DoOs08}.
We can, therefore, conclude that the homogeneous degree of any nonzero SL-invariant 
polynomial is $h\sum_{j=0}^{L-1}z_{j}$ for some integer $h$ and that the 
polynomial contains the monomial $(\prod_{j=0}^{L-1}\psi_{j}^{z_j})^h$.

\subsection{Irreducible a-balanced states and SU-invariants}
\label{secs4}

We have established that a partial spin flip operation mapping a c-balanced state 
$|\psi\rangle$ to an a-balanced state $|\tilde{\psi}\rangle$ induces a mapping 
between a homogeneous polynomial SL invariant that detects the c-balanced state 
and an invariant function that detects the a-balanced state. Furthermore, we have 
reviewed the multiplicative scaling behaviour of SL invariants on 
irreducible c-balanced states.

Now we address the question of the multiplicative scaling behaviour 
of the invariant $\tilde{P}$ for the case where
the map $P\to{\tilde{P}}$ is induced by a partial spin flip of some 
terms of an irreducible state.
In other words, we assume that $|\tilde{\psi}\rangle$ is irreducible 
and investigate how $\tilde{P}$ scales when $|\tilde{\psi}\rangle$ 
is multiplied by a factor $\lambda\in{\mathbb{C}\backslash\{0\}}$.
Let us assume that we have constructed the irreducible a-balanced state 
$|\tilde\psi\rangle$ from an irreducible c-balanced state by applying the 
partial spin flip operation to a subset $s$ of the $L$ terms that it has
(see Eq.~\eqref{tva}). 

\begin{Theorem} 
If $|{\psi}\rangle$ is an irreducible c-balanced state detected by a 
polynomial SL-invariant $P$ of homogeneous bidegree $(h\sum{z_{j}},0)$, 
and $|{\tilde{\psi}}\rangle$ is the irreducible a-balanced state constructed 
from $|{\psi}\rangle$ by spin flipping a subset $s$ of the terms, 
then the SU invariant $\tilde{P}$, constructed from $P$ by the induced mapping, 
has homogeneous bidegree $(h\sum_{j\notin{s}}{z_{j}},h\sum_{j\in{s}}z_{j})$.
\end{Theorem}
\begin{since}
Consider an irreducible a-balanced state
$|\tilde{\psi}\rangle=|\phi\rangle+|\tilde{\theta}\rangle$ 
and multiply it with a constant $\lambda$:
$|\tilde{\psi}\rangle_\lambda = \lambda \ket{\phi} +  \lambda |\tilde{\theta}\rangle$.
A spin-flip on the part $s$ (the $\tilde{\theta}$ part)
of the rescaled state gives
$|\tilde{\psi}\rangle_{\lambda}\to|\psi\rangle_{\lambda}=\lambda|\phi\rangle
+\lambda^*|\theta\rangle$. 
We immediately extract from \eqref{pa} that the SU invariant $\tilde{P}$ satisfies
$\tilde{P}(\lambda|\tilde{\psi}\rangle)=
\lambda^{h\sum_{j\not\in s} z_j}\lambda^{*h\sum_{j\in s}z_j}P(\ket{\psi})=
\lambda^{h\sum_{j\not\in s} z_j}\lambda^{*h\sum_{j\in s}z_j}\tilde{P}(|\tilde{\psi}\rangle)$.
We therefore have that $\tilde{P}$ is an $SU$ invariant of bidegree 
$(h\sum_{j\notin{s}}{z_{j}},h\sum_{j\in{s}}z_{j})$.
\end{since}

\begin{Corollary} 
Let $|\psi\rangle$ be an irreducible a-balanced state with associated integers 
$\{z_{0},z_{1},\dots,z_{L-1}\}$ that satisfy $\sum_{j=0}^{L-1}z_{j}\neq{0}$. 
Then, $|\psi\rangle$ is detected by an SU(2) invariant of bidegree $(d_{1},d_{2})$, 
where $d_{1}\neq{d_{2}}$, and this implies that $|\psi\rangle$ is a-balanced in every basis.
\end{Corollary}

Consider again the irreducible $c$-state $|\psi\rangle$ and the irreducible 
$a$-balanced state $|\tilde{\psi}\rangle$ produced by a spin flip operation on a subset 
$s$ of the terms as given by Eq. (\ref{tva}). From Eq. (\ref{pa}), we can easily find 
that the form of the part of $\tilde{P}$ that evaluates to a nonzero value, 
$\tilde{P}_{A}$, is

\begin{eqnarray}
\tilde{P}_{A}=i^{(\sum_{j\in{s}}\sum_{k=1}^{q}{A}_{kj})}
\prod_{j\notin{s}}\tilde{\psi}_{j}^{hz_{j}}\prod_{j\in{s}}\tilde{\psi}_{j}^{*hz_{j}}.
\end{eqnarray}
Every local SU invariant that detects states in the $A$ class of 
$|\tilde{\psi}\rangle$ contains a monomial of this form.

In \cite{OS09} it was shown that every irreducible c-balanced state 
is detected by a local SL(2) invariant polynomial. 
In a similar way, every irreducible a-balanced 
state is detected by a local SU(2) invariant polynomial.
Furthermore, in \cite{Johansson12} it was pointed out that if a state 
is detected by by polynomial local SU(2) invariants of bidegree 
$(d_{1},d_{2})$ such that $d_{1}\neq{d_{2}}$, then it exhibits topological phases.
As found here, every state that is irreducibly 
a-balanced and satisfies $\sum_{j}z_{j}\neq{0}$ is of this kind. 

The states which are irreducibly a-balanced but not c-balanced 
belong to SLOCC-zero classes, that is, they are not detected by 
any SL-invariant. In other words, they constitute the SL-null cone.
Therefore, the above observation implies that SLOCC-zero classes can 
be split into states that exhibit topological phases and those that do not.

\subsection{Derived irreducible states, invariants, and topological phases}
\label{secs5}

We now elaborate on how the irreducible a-balanced states can be constructed from a given irreducible c-balanced state. 
Methods to construct irreducible c-balanced states were discussed 
in \cite{OS09,Johansson12}.
Consider that we have an irreducible c-balanced state with an associated 
multiset of integers $\{z_{0},z_{1},\dots,z_{L-1}\}$ that is detected by 
an invariant polynomial $P$ of bidegree $(h\sum_{j=0}^{L-1}z_{j},0)$. 
By spin flipping different submultisets $s$ such that the associated 
submultisets of the integers $\{z_{0},z_{1},\dots,z_{L-1}\}$ satisfy  
$\sum_{j\in{s}}z_{j}\neq{\sum_{j\notin{s}}z_{j}}$, 
we can produce different irreducible a-balanced states that feature 
topological phases. Two different partial spin flips 
corresponding to different submultisets $s_{1}$ and $s_{2}$ may map the state 
to the same $A$ class, but if 
$\sum_{j\in{s_{1}}}z_{j}\neq\sum_{j\in{s_{2}}}z_{j}$ (or if one identifies with the 
complex conjugate classes: $\sum_{j\in{s_{1}}}z_{j}\neq{\sum_{j\notin{s_{2}}}z_{j}}$), then 
the $A$ classes are certainly distinct.

If we collect all states derived from $|\psi\rangle$, such that the set of 
spin flipped terms $s$ satisfies $\sum_{j\in{s}}z_{j}<{\sum_{j\notin{s}}z_{j}}$, 
these are representatives of each $A$ class that can be constructed in this way 
from $|\psi\rangle$. To each $A$ class, there is a corresponding invariant of bidegree 
$(h{\sum_{j\notin{s}}z_{j}},h\sum_{j\in{s}}z_{j})$.
Moreover, since the topological phases for 
$|{\psi}\rangle$ are multiples of $\frac{2\pi}{\sum_{j}{z_{j}}}$, we deduce that the topological phases for a state $|\tilde{\psi}\rangle$ constructed 
by spin flipping a set $s$ are multiples of $\chi$ where

\begin{eqnarray}
\chi=\frac{2\pi}{(|\sum_{j\notin{s}}{z_{j}}|-|\sum_{j\in{s}}{z_{j}}|)}.
\end{eqnarray}

Let us consider the case when the submultiset $s$ satisfies 
$\sum_{j\in{s}}z_{j}=\sum_{j\notin{s}}z_{j}$, that is, when it defines an equal partition 
of $\{z_{0},z_{1},\dots,z_{L-1}\}$. In each row of the matrix $A_{jk}$, the $1$s and $-1$s 
also define an equal partition of $\{z_{0},z_{1},\dots,z_{L-1}\}$. 
If $s$ defines an equal partition that also corresponds to a row of $A_{jk}$, 
the partial spin flip produces a row of only $1$s or only $-1$s. 
In this case, the qubit corresponding to this row is in a tensor product 
with the remaining qubits.
However, if there are more equal partitions of the $\{z_{0},z_{1},\dots,z_{L-1}\}$ 
than $A_{jk}$ has rows, it is possible to select an $s$ such that a non-product 
state results. 
In this case, one has produced an entangled state that can only be detected by 
polynomials that are invariant under the full 
${\rm{U(2})}^{\rm{\otimes{q}}}$ group and that does not have topological phases. 
These states are thus natural extensions of the $W$ state.

Returning to the case of three qubits we can see that a partial spin flip of 
a single term of an irreducible state in the $X$ class will produces a state 
in the $W'$ class. A spin flip of two terms produces a state where one 
qubit is in a tensor product with a possibly entangled two-qubit state.

As a further example, consider the five-qubit irreducible c-balanced state from
\cite{OS09}
\beq\label{in}
\ket{11111}+\ket{11000}+\ket{10110} + \ket{01000} + \ket{00101}  + \ket{00011}
\; ,
\eeq
which, by a spin flip on the basis state $s=\{1\}$, is transformed into the
irreducible a-balanced state
\beq
\ket{00000}+\ket{11000}+\ket{10110} + \ket{01000} + \ket{00101}  + \ket{00011}
\eeq
with non-zero invariant of bidegree $(5,1)$, then by spin flip on the basis state $s=\{2\}$
into an irreducible a-balanced state,
\beq
\ket{00000}+\ket{00111}+\ket{10110} + \ket{01000} + \ket{00101}  + \ket{00011}
\eeq
with non-zero invariant of bidegree $(4,2)$, and finally by a spin flip on the basis state
$s=\{5\}$ into a state with a non-zero invariant of bidegree $(3,3)$,
\beq\label{out}
\ket{00000}+\ket{00111}+\ket{10110} + \ket{01000} + \ket{11010}  + \ket{00011}\; .
\eeq
This final state, which comes out of the initial irreducible c-balanced state~\eqref{in}
by a spin flip on the part $s=\{1,2,5\}$, is a state that is only detected by 
${\rm{U(2})}^{\rm{\otimes{5}}}$ invariants, similarly to the 
$\ket{W}$ states, which however are completely unbalanced.

We can, in fact, go even further than spin flipping submultisets $s$, and instead split a 
basis product state vector into two parts, followed by a spin flip on only one of the parts. 
However, the resulting state is then a GHZ state plus an unbalanced state.
In order to give an example with the splitting of a basis product state vector, 
we take the $X$ state from~\cite{OS05}
\beq
\sqrt{2}\ket{1111}+W \leftrightarrow A=\Matrix{cccccc}{
1&1&1&0&0&0\\
1&1&0&1&0&0\\
1&1&0&0&1&0\\
1&1&0&0&0&1}
\; .
\eeq
This state is mapped by a partial spin flip on $s=\{1,3,4\}$ into
\beqa
\ket{1111}+\ket{0000}&&\\
\ket{0111}+\ket{1011}&+&\ket{0010}+\ket{0001}\; .
\eeqa
The splitting into a GHZ (c-balanced) state and the $W$-like (unbalanced) state, 
\beq
\ket{W-like}=\ket{0111}+\ket{1011}+\ket{0010}+\ket{0001}
\eeq 
is easily seen.

The number of different types of irreducible c-balancedness for $q$ qubits increases rapidly with $q$. The problem of finding these different types can be rephrased as a combinatorial problem which can be solved algorithmically. However, the algorithmic search constructed in Ref. \cite{Johansson12} already becomes computationally expensive for $q>7$. Constructing the invariants corresponding to the irreducible a-balanced or c-balanced states is also a demanding task and, in general, the invariants of qubit systems have only been studied for up to five qubits \cite{Luque05,DoOs08}.

\section{Induced classification}
\label{ind}

Any qualitative entanglement property can be used to classify entangled states. 
Local SU(2) interconvertibility already groups the states into classes, since 
belonging to the same orbit of some group is an equivalence relation. 
A complete generating set of invariants distinguishes all the orbits of the underlying group.
In this way, a finite number of invariants produces an infinite number of classes. 
However, for some purposes, it may be useful to classify states based on some property 
of interest that yields a finite number of classes. Here, we reflect upon how balancedness 
can be used to construct such a classification.

One approach is to classify states based on the balancedness of their minimal form. 
However, states may have several minimal forms with different balancedness, thus making 
the assignment non unique.  
Moreover, the minimal forms do not always give us the full picture.
One such example is the intersection of the GHZ-class and the $X$ class in Sec. \ref{3bit}. 
While the states in the local SU orbit of the GHZ state have the GHZ state itself 
as a minimal form, they can also be put on the form of the $X$ state.

Any SL- or SU-invariant polynomial that is not invariant under U(1) transformations 
detects only balanced parts of states. However, the choice of homogeneous generators 
for the polynomial algebra of invariants is typically not unique. Given a generator, 
products of other generators can often be added without changing the bidegree of the 
generator.
As we saw in Sec. \ref{gen}, for every irreducible state, there is at least one 
invariant polynomial containing a monomial term that precisely captures the balancedness 
of the state. This allows us to choose generators of the polynomial algebra such that 
a generator detects only a given type of irreducibly balanced states while not detecting any other 
types.  
This connection between the different kinds of c- or a-balancedness and invariant 
polynomials is a direct result of SL or SU invariance, respectively. 
We therefore use the invariants chosen in this way as a starting point for a classification.  

Because our interest is in a qualitative classification, we consider only which 
invariants out of a generating set take nonzero values, rather than what 
their precise values are.   
Assume that a complete generating set for the invariants is known.
Then we put two states in the same class, if they are detected
by the same set of invariants. 
This is an equivalence relation (self-similarity and transitivity) and groups 
the states in finitely many classes for finitely many qubits.
Given that we have chosen the generators, as outlined above, this also captures 
the balancedness of the states.

Here, it is worth mentioning that some carefully chosen elements of the zero class 
of the underlying symmetry group can be added without changing the class. 
That is, if a state vector from the zero class can be added to a given vector 
without creating new balancedness in the resulting state, this careful adding does 
not change the class. 
This can be elements out of the $(n,n)$ class plus 
the unbalanced class for classifications with respect to SU(2), 
or the $(2n-p,p)$ class, $p\not\in \{0,2n\}$, 
classifying with respect to the group SL(2).

Beyond the connection to topological phases, it is unclear to us what physical sense 
a classification of the kind given here may carry but it is a very natural one that 
relies on the SU invariance (or SL invariance) of entanglement properties, 
and we consider further analysis worthwhile.

\subsubsection{Examples on how the classification works}

As an example, we can first consider the connection between different types of c-balancedness 
and polynomial local SL invariants in the case of four qubits which was previously 
investigated in \cite{OS09,Viehmann}. 
The generating set of polynomial local SL invariants contains four generators of 
polynomial degree 2, 4, 4, and 6, respectively~\cite{Luque02}.
In Ref.~\cite{OS05}, irreducible c-balanced states that represent up to seven different 
types (invariant under qubit permutations) of genuine four-qubit entanglement were 
identified~\cite{Ost13}. One is the four-qubit GHZ state,

\begin{equation}|0000\rangle+|1111\rangle,\end{equation}
which is a representative of the only type of genuinely multipartite entanglement that is 
detected by the generator of degree 2. The second is the cluster state

\begin{equation}\label{telescope}
|1111\rangle+|1100\rangle+|0010\rangle+|0001\rangle,
\end{equation}
which is local unitary equivalent to $|0000\rangle+|0011\rangle+|1100\rangle-|1111\rangle$, 
and the states related to the state in Eq. (\ref{telescope}) by permutation of the qubits. 
These are only detected by the two degree 4 generators after functional dependencies 
of the generators have been removed. The last state is the four-qubit $X$ state

\begin{equation}
|1111\rangle+|1000\rangle+|0100\rangle+|0010\rangle+|0001\rangle,
\end{equation}
which is only detected by the degree 6 generator.
As found in \cite{Johansson12}, these states are also distinguished by their respective 
topological phases. The four-qubit GHZ state has only the topological phase $\pi$, 
while the four-qubit cluster state has a topological phase $\frac{\pi}{2}$ and the 
four-qubit $X$ state has a phase $\frac{\pi}{3}$. 

In Ref. \cite{Viehmann}, a classification scheme was suggested for four-qubit entangled states 
where SLOCC-equivalence classes were grouped into families in terms of "tangle patterns".
The tangle patterns are defined in terms of a generating set of polynomial invariants, and 
the highest degree of a generator that detects a state determines which family it belongs to.  
In this way, the states are sorted into a hierarchy of four families. 
The four families are precisely the c-unbalanced states, and three families 
corresponding to the four generators of degree 2, 4, and 6, respectively.
The entanglement types of the three genuine entangled families have been named after 
representative states of the respective family that
is detected only by its particular highest degree generator. These are the states identified 
in \cite{OS05} and therefore the types are called $X$ type, cluster type and 
GHZ type \cite{Viehmann}. 
In addition to this, there is the $W$ type family which is not detected by 
$SL$ invariant polynomials.
\\
We note that the families are distinguished by topological phases under  
cyclic local $SL$-evolution just as are their representative states. 
Only the $X$ type family contains states with topological phase $\frac{\pi}{3}$ under cyclic 
local $SL$ evolution. Only the cluster-type family contains states with topological phase 
$\frac{\pi}{2}$. All states in the GHZ type family have the phase $\pi$ and the states of 
the $W$ type have no topological phases under cyclic local SL evolution.

In contrast to this, our classification scheme gives a slightly different answer 
and is more fine grained. From the chosen invariants 
$({\cal A},{{\cal B}^{I}},{{\cal B}^{II}},{\cal C} )$ \cite{Viehmann} of polynomial
degrees $(2,4,4,6)$, we can construct the invariants 
$({\cal A},{{\cal B}^{I}-{\cal A}^{2}},{{\cal B}^{II}-{\cal A}^{2}},{\cal C +A}^3)$. 
These satisfy the property that each invariant detect only one kind of 
irreducible balancedness.

Using these invariants, we give a 4-tuple $(a,b_1,b_2,c)$ in $\ZZ_2^4$.
The class $(1,0,0,0)$ corresponds to the GHZ-type entanglement, while $(0,1,0,0)$ and 
$(0,0,1,0)$ correspond to the two different kinds of irreducibly balanced states of 
length 4, i.e., the two different kinds of cluster type states related by permutations 
of the second and third qubit. The irreducibly balanced states of length 6 i.e. states 
with $X$-type entanglement belong to the class $(0,0,0,1)$. 
The two different kinds of biseparable states of the type 
$|\phi_{1}\rangle\otimes|\phi_{2}\rangle$ where each $|\phi_{i}\rangle$ is an entangled 
state of two qubits are found in the two classes $(1,1,0,1)$ and $(1,0,1,1)$. 
These two classes also contain the reducibly balanced states of length 4 that are not biseparable. Almost every state except a set of zero measure belongs 
to the class $(1,1,1,1)$ containing all the different kinds of c-balancedness. 
$(0,0,0,0)$ is the class of states that are not c-balanced. 
They will be treated in what follows.
This classification gives us at most 15 classes of genuine four-party entanglement.

When looking at a classification of states based on their a-balancedness and their 
topological phases under local SU(2) evolution, we make the more fine grained 
division of local SU(2) orbits into families based 
on which local SU(2) invariants of bidegree $(d_{1},d_{2})$, 
where $d_{1}\neq{d_{2}}$, detect them.
We note that such a classification scheme captures the structure on the set of entangled 
three qubit states that was described in Sec. \ref{3bit}. Here the GHZ SLOCC-class is 
subdivided into the $X$ family detected only by the threetangle and the rest of 
the GHZ SLOCC class detected by both the threetangle and $\tau_{3,1}$. 
The $W$ family is divided into the $W'$ family detected by $\tau_{3,1}$ 
with the $W'$-state as a representative 
and the unbalanced states with the $W$ state as a representative.
In particular, we see that such a scheme divides the W SLOCC-class of states, i.e. the states 
that are not c-balanced, into subfamilies based on a-balancedness through the associated local 
SU(2)-invariant polynomials.

As a further example of this, we can consider the four qubit states that are not c-balanced. 
For four qubits, the irreducible a-balanced states derived from the $X$ state are detected by 
invariants of bidegree (5,1) or of bidegree (4,2) and display topological phases 
$\frac{\pi}{2}$ or $\pi$ respectively. The irreducible a-balanced states that can be derived 
from the cluster state in Eq. (\ref{telescope}) are detected by polynomials of bidegree (3,1) and 
display the topological phase $\pi$. 
By considering the different combinations of these invariants that can detect a state, and treating states related by qubit permutations as equivalent, 
we get, at most, seven subfamilies. In addition to this, we have the family of a-unbalanced 
states that are only detected by invariants of the full local unitary group.  

As an example of the classification scheme involving c-balanced states, we can consider the 
local SU orbits of the four-qubit $X$-type family from \cite{Viehmann}. 
These are all detected by the generator of bidegree (6,0) but can be further subdivided into subfamilies 
based on which of the invariants of bidegree (4,0), (3,1), (5,1), (4,2), and (2,0)  
detects the states. This gives, at most, a total of 32 subfamilies. 
Notably, only the subfamily detected 
by none of the polynomials other than that of bidegree (6,0) contains states with 
topological phase $\frac{\pi}{3}$, while the other subfamilies only contain states with phase $\pi$.

This classification scheme gives us a hierarchy of entanglement families based on the concept 
of balancedness. Furthermore, it is closely connected to the qualitative feature of 
topological phases displayed by states in the respective family. Further investigation of this concept
would be highly desirable.

\section{Conclusions}\label{concl}
 
Some examples of irreducible a-balanced states with topological phases are known from 
\cite{Johansson12}, but in this work we have demonstrated
that topological phases are a feature of all irreducibly 
a-balanced states for which the associated set of integers has a nonzero sum. We have also shown 
that every such state is detected by a local SU-invariant polynomial of bidegree 
$(d_1,d_2)$, where $d_1\neq{d_2}$. 
It distinguishes these states from the unbalanced states that do not have any
topological phase and which are only detected by invariants of bidegree $(d_1,d_1)$.

We have shown this by introducing a partial spin-flip, which maps between states that are
c-balanced and those which are only a-balanced. The partial spin-flip map also induces a map between
invariants, such that from an initial $SL(2)$ invariant the existence of $SU(2)$ invariants
that detect irreducible a-balanced states follows.
For three qubits, an invariant $\tau_{3,1}$ of bidegree $(3,1)$ detects 
the irreducible a-balanced state $|\tilde{W}\rangle$ with topological phase $\pi$  
but not the unbalanced $W$ state that has no topological phases.
The invariant $\tau_{3,1}$ also distinguishes between the genuinely threepartite entangled 
states with topological phase $\pi$ and those with topological phases $\frac{m\pi}{2}$ 
for integer $m$.
The remaining states are detected only by invariants of the group of local U(2), and thus are 
in the zero-class of both SL(2) and SU(2) invariants. The set of these states contains 
bipartite product states besides globally entangled states, which are $W$ like states. 
Which class a state belongs to can be clearly foreseen from the original (irreducible) 
c-balanced form of an SL(2)-invariant state. 

Furthermore, we have discussed how balancedness and, in particular, a-balancedness, 
can be used for the classification of entangled states. 
The connection to polynomial invariants as well as topological phases suggests 
that c-balancedness as well as a-balancedness are useful concepts for the description of multipartite entangled states.

\acknowledgments
M.J. and E.S. acknowledge support from the National Research Foundation and the Ministry of
Education (Singapore). 
M.E. acknowledges support from the Swedish Research Council (VR).
A. O. acknowledges financial support by the DFG within the SFB TR12.

\section{Appendix}

Let us now illustrate what a-balancedness means by showing that irreducible 
a-balanced states for which $\sum_{j}z_{j}\neq{0}$ always have an a-balanced part after local SU-transformations 
on a single qubit. 
To see this, let an SU transformation $U$ act on a single qubit of an a-balanced state 
$|\psi\rangle$ whose balancedness is represented by a matrix $A$. 
We can express the resulting state as a sum of two parts 
$U_{1}|\psi\rangle$ and $U_{2}|\psi\rangle$ corresponding to the action of the diagonal part 
$U_{1}=\frac{1}{2}(\id\trace U +\trace(U\sigma_{z})\sigma_{z})$ and the off-diagonal part 
$U_{2}=\frac{1}{2}(\trace(U\sigma_{x})\sigma_{x}+\trace(U\sigma_{y})\sigma_{y})$ of $U$ respectively.
Both parts $U_{1}|\psi\rangle$ and $U_{2}|\psi\rangle$ are a-balanced with the same set of 
integer numbers $z_1,\dots , z_L$ and their respective matrices are $A_{1}=A$, and $A_{2}$ 
related to $A$ by multiplication of a row by $-1$. 
If no pair of columns in $A_{1}$ is also found in $A_{2}$, the matrix of the full state 
contains the columns of both $A_{1}$ and $A_{2}$ and is then clearly balanced. 
However, if $A_{1}$ and $A_{2}$ has one or more pairs of columns in common it is possible to 
choose the SU transformation such that one or possibly several of the terms
corresponding to these columns cancel out. In this case, these columns do not 
appear in the matrix of the full state and therefore a-balancedness may be lost.

However, if the original state
 is irreducibly a-balanced ,i.e., if no proper subset of columns in $A$ are linearly dependent, and if $\sum_{j}z_j\neq{0}$, it follows that at most one pair of columns is common to $A_1$ and $A_2$. 
Even if a term corresponding to such a column is cancelled out the full state is still balanced.
To see this let $z_l$ and $z_{l'}$ be the integers that correspond to the columns belonging 
to the cancelled term.
By choosing $m_{1} z_l + m_{2} z_{l'}=0$, with $m_{1}$ and $m_{2}$ relatively prime, the integer 
corresponding to the cancelled term is zero.
The corresponding integers for the remaining state is 
$m_{1} (z_1,\dots , z_{l-1},z_{l+1},\dots,z_L)_{A_{1}},
 m_{2} (z_1, \dots , z_{l'-1},z_{l'+1},\dots,z_L)_{A_{2}}$
and thus the state is balanced also in the 
absence of this column, corresponding to the irreducible balancedness of $A_{1}$ and $A_{2}$ 
respectively.

As an example consider the irreducible a-balanced three-qubit state
\beq
\ket{W'}= \ket{000}+\ket{100}+\ket{010}+\ket{001}
\eeq
with integer numbers $(-1,1,1,1)$ describing its a-balancedness, and apply the Hadamard 
transformation on the first qubit. Here, the term $\ket{100}$ is cancelled out by the 
transformation and the resulting state is
 
\beqa
H_1 &=& \Matrix{cc}{1&\phantom{-}1\\1&-1}\\
H_1 \ket{W'}&=&\ket{000}+\ket{010}+\ket{110}+\ket{001}+\ket{101}.
\eeqa

Therefore, we find the integers corresponding to the matrices $A_{1}$ and $A_{2}$ by 
choosing $m_{1}=m_{2}=1$ 
\beq
A_{1}=\stackrel{\!\!\!\begin{array}{cccc}-1&\phantom{-}1&\phantom{-}1&\phantom{-}1\end{array}}{\Matrix{cccc}{-1&\phantom{-}1&-1&-1\\
-1&-1&\phantom{-}1&-1\\
-1&-1&-1&\phantom{-}1}}\; ; \
A_{2}=\stackrel{\!\!\!\begin{array}{cccc}-1&\phantom{-}1&\phantom{-}1&\phantom{-}1\end{array} }{ \Matrix{cccc}{\phantom{-}1&-1&\phantom{-}1&\phantom{-}1\\-1&-1&\phantom{-}1&-1\\
-1&-1&-1&\phantom{-}1}}
\ .
\eeq
The matrix and integers of the full state are
\beq
A_{H_{1}|\psi\rangle}=\stackrel{\!\!\!\begin{array}{ccccc}\phantom{-}0&\phantom{-}1&\phantom{-}1&\phantom{-}1&\phantom{-}1\end{array}}{\Matrix{ccccc}{-1&-1&\phantom{-}1&-1&\phantom{-}1\\
-1&\phantom{-}1&\phantom{-}1&-1&-1\\
-1&-1&-1&\phantom{-}1&\phantom{-}1}}\; .
\eeq
As in the example, one or several of the integer numbers attributed to product basis states can become zero but
the remaining components are (a- or c-) balanced. 
In particular, the resulting state cannot be a product state.

A second example is the four-qubit state
\beq
\ket{W'}= \ket{0000}+\ket{1000}+\ket{0100}+\ket{0010}+\ket{0001}
\eeq
with corresponding integers $(-2,1,1,1,1)$.
Using the same transformation on the first qubit, we get the matrices
\beq
A_{1}=\stackrel{\!\!\!\begin{array}{ccccc}-2&\phantom{-}1&\phantom{-}1&\phantom{-}1&\phantom{-}1\end{array}}{\Matrix{ccccc}{-1&\phantom{-}1&-1&-1&-1\\
-1&-1&\phantom{-}1&-1&-1\\
-1&-1&-1&\phantom{-}1&-1\\
-1&-1&-1&-1&\phantom{-}1}}\; ;\ 
A_{2}=\stackrel{\!\!\!\begin{array}{ccccc}-2&\phantom{-}1&\phantom{-}1&\phantom{-}1&\phantom{-}1\end{array}}{\Matrix{ccccc}{\phantom{-}1&-1&\phantom{-}1&\phantom{-}1&\phantom{-}1\\-1&-1&\phantom{-}1&-1&-1\\
-1&-1&-1&\phantom{-}1&-1\\
-1&-1&-1&-1&\phantom{-}1}}
\ .
\eeq
Here, $\ket{1000}$ is cancelled ($m_1=2,m_2=1$) and we have
\beq
H_1 \ket{W'}=\ket{0000}+\ket{0100}+\ket{0010}+\ket{0001}+\ket{1100}+\ket{1010}+\ket{1001}
\eeq
with corresponding integers $(-3,2,2,2,1,1,1)$.

\end{document}